\documentclass[
showpacs,twocolumn,superscriptaddress,
floatfix,amsmath,amsfonts,amssymb,
10pt
]{revtex4}

\usepackage{graphicx}\graphicspath{{figs/}}
\usepackage{comment}

\begin{document}
  \title{Generalised Atmospheric Rosenbluth Methods (GARM)}
  \author{A.\ Rechnitzer} \email{andrewr@math.ubc.ca}
  \affiliation{Department of Mathematics, University of British Columbia,
  Canada}
  \author{E.\ J.\ Janse van Rensburg} \email{rensburg@yorku.ca}
  \affiliation{Department of Mathematics and Statistics, York University,
  Canada}
\date{\today}
\begin{abstract}
  We show that the classical Rosenbluth method for sampling self-avoiding
walks \cite{hammersley1954,rosenbluth1955} can be extended to a general
algorithm for sampling many families of objects, including self-avoiding
polygons. The implementation relies on an elementary move which is a
generalisation of kinetic growth; rather than only appending edges to the
endpoint, edges may be inserted at any vertex providing the resulting objects
still lie within the same family. We implement this method using pruning and
enrichment~\cite{grassberger1997} to sample self-avoiding walks and polygons.
The algorithm can be further extended by mixing it with length-preserving moves,
such pivots and crank-shaft moves.
\end{abstract}
\maketitle
Monte Carlo simulations of self-avoiding walks (SAWs) and self-avoiding polygons
(SAPs) on regular lattices are a major tool for the study of polymer statistics
\cite{degennes1979}. While kinetic growth algorithms \cite{hammersley1954,
rosenbluth1955} have been used to sample SAWs to great success, it is
unclear how they might be applied to SAPs. SAPs are models of ring polymers,
plasmids (mitochondrial DNA) \cite{rybenkov1993, gee1997, marcone2007} and
appear in the zero-component limit of the $N$-vector model \cite{symanzik1969}.
In this paper, we generalise a growth algorithm for SAWs and show how it may be
used to sample SAPs and other objects.

The Rosenbluth method for sampling self-avoiding walks (SAWs) is a classical
algorithm dating back to the 1950s \cite{hammersley1954, rosenbluth1955}. This
method found new application with the development of a pruned and enriched
implementation, called PERM, due to Grassberger \cite{grassberger1997}. This was
in turn further extended using flat-histogram and multicanonical methods
\cite{bachmann2003,prellberg2004}. These new algorithms have found many
applications in the modelling of polymers (eg.\cite{causo2000, krawczyk2007,
hsu2007}).

The Rosenbluth method samples SAWs by growing conformations from
a single vertex and iteratively adding edges from the final vertex to
its unoccupied nearest neighbours. Let $\omega$ be a SAW of $n$ edges and let
$m_k$ be the number of unoccupied nearest neighbour vertices of its endpoint
when it is truncated after $k$ edges (after the $k^\mathrm{th}$ iteration). The
probability of sampling $\omega$ is given by
\begin{align}
  \Pr(\omega) &= \prod_{k=1}^{n} m_{k-1}^{-1}.
\end{align}
If $m_k=0$ the conformation is trapped and a new conformation is started. This
attrition makes it difficult to sample long SAWs and sampling by this method is
not uniform. The sample bias can be removed by weighting each SAW by $W(\omega)
= \Pr(\omega)^{-1}$. The mean weight of all SAWs of length $n$ is
\begin{align}
  \sum_{|\omega|=n} W(\omega) \Pr(\omega) &= c_n
\end{align}
where $c_n$ is the number of SAWs of length $n$. Thus one can estimate
the number of conformations by computing the average weight of sampled
SAWs. Sample averages of observables are computed by taking weighted
 averages.
 
The original Rosenbluth algorithm converges very slowly because of attrition
and the large variance in weights of longer SAWs. Grassberger's pruned and
enriched implementation \cite{grassberger1997} constituted a major advance that
allows the algorithm to reduce these effects. This significantly improves its
efficiency and applicability.

In this paper we demonstrate how the Rosenbluth algorithm can be generalised;
SAWs can be sampled by inserting edges at any vertex rather than only at the
endpoint. This idea may be extended to a growth algorithm for SAPs and other
lattice objects. We also show how to combine it with length preserving
moves such as pivots.

\vspace{-2ex}
\section*{Atmospheres}
\begin{figure}[h]
  \vspace{-1ex}
  \begin{center}
  \includegraphics[height=3.5cm]{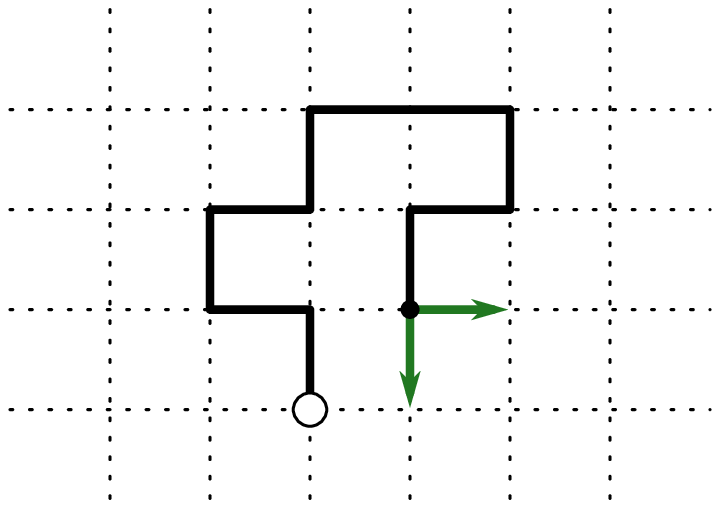}
  \end{center}
  \vspace{-4.5ex}
  \caption{A SAW and its two positive endpoint atmospheric edges.}
  \label{fig endpoint}
\end{figure}

\vspace{-2ex}
Let $\omega$ be a SAW on the square lattice, starting from the
origin. A \emph{positive endpoint atmospheric edge} of $\omega$ is an edge on
the lattice that can be appended to the last vertex of $\omega$ to extend its
length by one while respecting self-avoidance (see Figure~\ref{fig endpoint}).
The size of the positive atmosphere of $\omega$, $a_+(\omega)$, is the number
of positive atmospheric edges. We abuse our notation and use $a_+$ to denote
both the set of positive atmospheric edges and its size. Adding positive
endpoint atmospheric edges increases the length of the SAW.

\begin{figure}[h!]
\vspace{-5ex}
\begin{center}
  \includegraphics[height=3.5cm]{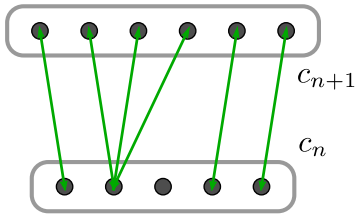}
\end{center}
\vspace{-4.5ex}
\caption{A schematic picture of the linkages between $c_n$ and~$c_{n+1}$. There
is one conformation with empty positive atmosphere.}
\label{fig linkage}
\end{figure}

Appending a positive atmospheric edge to $\omega$ to obtain~$\omega'$ creates
a linkage $(\omega,\omega')$ (see Figure~\ref{fig linkage}). Deleting the
last edge from $\omega'$ gives $\omega$; we define the \emph{negative endpoint
atmosphere} of $\omega'$ to be this edge. We denote the size of the negative
atmosphere of $\omega'$ by $a_-(\omega')$ and we again abuse notation by
also using this symbol to denote the set of such edges. In the present context
$a_-(\omega') \equiv 1$, but below we consider more general positive and
negative atmospheres.

By counting the number of linkages, we see that
\begin{align}
\# \mbox{linkages} &= \sum_{\omega} a_+(\omega) = \sum_{\omega'} a_-(\omega')
= c_{n+1}.
\end{align}
This implies that
\begin{align}
\label{eqn atm ratio}
\frac{\langle a_+ \rangle_n }{\langle a_- \rangle_{n+1} } &=
\frac{c_{n+1}}{c_n}
\end{align}
where $a_-(\omega') \equiv 1$ and the averages are taken over the uniform
distribution.

This observation can be used to estimate growth constants and
free-energies of SAWs and bond trees \cite{rechnitzer2002, vanrensburg2003,
vanrensburg2004}. We extend these definitions and show how they lead to a
significant generalisation of the Rosenbluth algorithm.

\begin{figure}[h]
  \vspace{-1ex}
  \begin{center}
  \includegraphics[height=3.0cm]{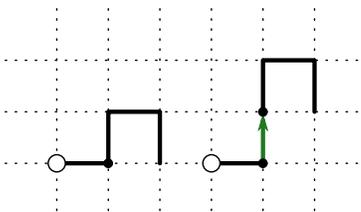}
  \end{center}
  \vspace{-4.5ex}
  \caption{The SAW on the right is obtained from the SAW on the left by
  inserting a north edge at the black vertex. This is one of its eleven
  positive atmospheric edges. It has three negative atmospheric edges.}
\label{fig genatm}
\end{figure}

Define the positive atmosphere, $a_+(\omega)$, to be the number of ways that an
edge can be  inserted into $\omega$ at any of its vertices so that a SAW is
obtained (see Figure~\ref{fig genatm}). Note that there are SAWs with empty
positive atmosphere.

\begin{figure}[h]
  \vspace{-1ex}
\begin{center}
  \includegraphics[height=5.5cm]{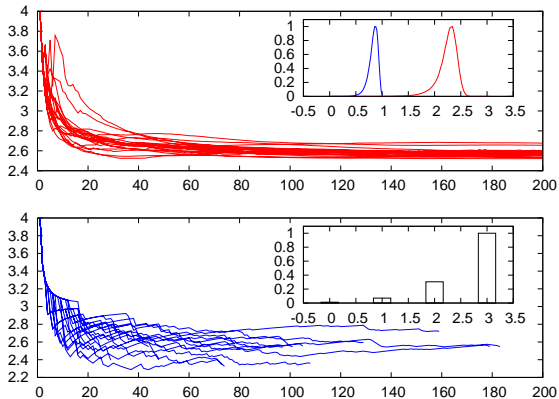}
\end{center}
  \vspace{-4ex}
\caption{Typical evolution of $W(\varphi)^{1/n}$ for both Rosenbluth and GARM
sampling. Shown are 20 samples from each. While GARM does suffer from
attrition, SAWs sampled by Rosenbluth have larger variance and a higher rate of
attrition. The insets show the distributions of positive and negative
atmospheres per vertex (top) and endpoint atmospheres (bottom); the peak
heights have been normalised to 1.}
\label{fig saw seq}
\end{figure}

Inserting a positive atmospheric edge into $\omega$ results
in a new SAW, $\omega'$. The negative atmosphere, $a_-(\omega')$, is the number
of ways that an edge can be deleted to obtain a SAW. Linkages are
created as above and equation~\eqref{eqn atm ratio} holds by the same
arguments. Simulations show that the distribution of atmospheres are
narrowly peaked (see Figure~\ref{fig saw seq}).

\vspace{-2ex}
\section*{Generalised atmospheric Rosenbluth method}
\vspace{-2ex}
The generalised atmospheres, $a_\pm$, can be used to define a generalised
Rosenbluth method for sampling SAWs. The algorithm starts with a single
vertex, $\varphi_0$, and grows a sequence of SAWs, $\varphi = \varphi_0, \dots,
\varphi_n$, by inserting a positive  atmospheric edge at each iteration. The
conformation  $\varphi_{k+1}$ is obtained from $\varphi_k$ by inserting an edge
chosen  uniformly from the available positive atmospheric edges,
$a_+(\varphi_k)$. We call this the Generalised Atmospheric Rosenbluth Method
(GARM). This method generalises the percolation based algorithms for trees in
\cite{care2000,hsu2005}.

A \emph{sequence} of $n+1$ SAWs, $\varphi = \varphi_0,\dots, \varphi_n$, is
obtained after $n$ iterations with probability
\begin{align}
  \Pr(\varphi\,|\,\varphi_0) &= \prod_{k=1}^{n} a_+(\varphi_{k-1})^{-1}.
\end{align}
Since a given conformation can be obtained in several different ways, this is
\emph{not} the probability of obtaining the last SAW in the sequence. As such we
give a weight to the sequence of SAWS, not only to compensate for the
non-uniform sampling probability, but also to take into account this
degeneracy. The weight of a sequence of SAWs, $\varphi = \varphi_0, \dots,
\varphi_n$ is
\begin{align}
W(\varphi)
&= \prod_{k=1}^{n} \frac{a_+(\varphi_{k-1})}{a_-(\varphi_{k})}
\end{align}
if $n \geq 1$ and $W(\varphi)=1$ if $n=0$.

The mean weight of sequences of length $n+1$ is
\begin{align}
\langle W \rangle_n &= \sum_{\varphi} W(\varphi) \Pr(\varphi \,|\, \varphi_0)
= c_n.
\end{align}
To see this, consider all the sequences that end in a particular SAW~$\tau$ of
length $n$. It suffices to show that
\begin{align}
\label{eqn phi sum}
1 &= \sum_{\varphi \to \tau} W(\varphi) \Pr(\varphi \,|\,\varphi_0)
= \sum_{\varphi \to \tau} \prod_{k=1}^{n} a_-(\varphi_{k})^{-1}
\end{align}
where the sums are over all sequences that end in $\tau$. Note that one must
consider all the possible choices of atmospheric edges, so that there are $n!$
sequences that end in the SAW made up of $n$ east edges.

We reinterpret the product of negative atmospheres as the probability of
returning to the single vertex under the following process. Starting at
$\varphi_n$, we delete negative atmospheric edges from $\varphi_k$ to obtain
$\varphi_{k-1}$ iteratively. The probability of realising $\varphi_0$ along the
sequence $\varphi$ is
\begin{align}
  \Pr(\varphi_0 | \varphi)
  &= \prod_{k=1}^{n} \Pr(\varphi_{k-1} \,|\, \varphi_k)
  = \prod_{k=1}^{n} a_-(\varphi_k)^{-1}.
\end{align}
Since all sequences that end at $\tau$ must return to $\varphi_0$ by this
process, summing over $\varphi$ gives equation~\eqref{eqn phi sum} as required.

The above proof becomes trivial in the  case of the endpoint atmosphere since
$a_- \equiv 1$  and each SAW is obtained in exactly one way. The proof breaks
down in models in which a given conformation cannot be reached by inserting
positive atmospheric edges.

\begin{figure}[b]
  \vspace{-1ex}
\begin{center}
  \includegraphics[height=4.0cm]{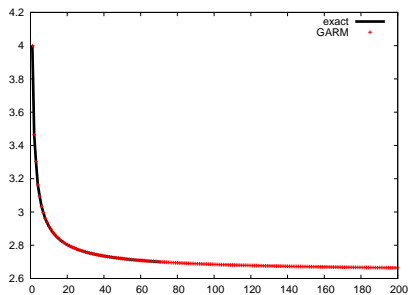}
\end{center}
  \vspace{-4ex}
\caption{A plot of $c_n^{1/n}$ as estimated from GARM data up to length
200. Exact enumeration data up to length 71 taken from \cite{jensen2004} is
shown for comparison.}
\label{fig saw2d data}
\end{figure}
  
In Figure~\ref{fig saw2d data} we show that data obtained by a pruned
and enriched implementation of GARM for SAWs agrees with exact enumeration data
from \cite{jensen2004}. We generated SAWs of a maximum of 200 edges with
approximately $10^6$ trajectories consisting of a total of $2.5 \times 10^8$
samples. This took about an hour on a laptop computer. This algorithm for
the square-lattice generalises to SAWs on any graph with finite maximal
degree.

\vspace{-2ex}
\section*{Extensions to polygons, trees and animals}
\vspace{-2ex}
We now extend this algorithm to SAPs on the square
lattice. In a previous paper, we defined the positive atmosphere to be the
locations in which a  single edge can be replaced by a $\sqcap$ conformation
of three edges and the negative atmosphere was defined by the inverse of
this process \cite{vanrensburg2008}. This definition is insufficient for GARM
since there are many conformations that are not obtainable from the unit square;
for example the $2 \times 2$ square.

\begin{figure}[h]
  \vspace{-1ex}
\begin{center}
\includegraphics[height=3.0cm]{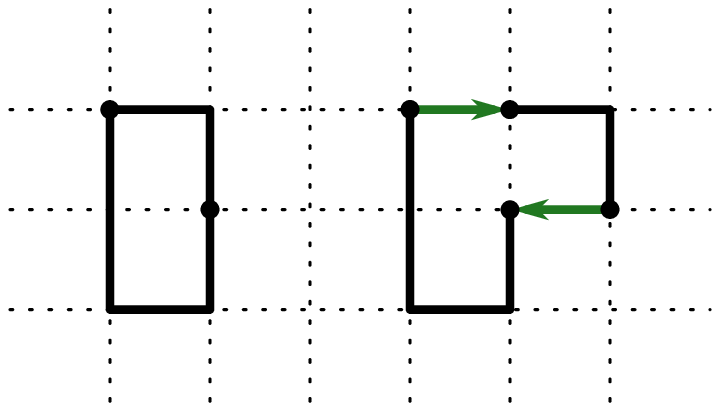}
\end{center}
\vspace{-4.5ex}
\caption{A SAP and the insertion of a pair of anti-parallel edges. This SAP has
positive atmosphere of 21 and negative atmosphere of 4.}
\label{fig sap atm}
\end{figure}

We generalise the notion of positive atmospheres of SAPs by considering all the 
pairs of vertices at which anti-parallel edges may be inserted to obtain a
longer SAP. The negative atmosphere is defined by finding all pairs of edges
that may be removed to obtain a SAP. See Figure~\ref{fig sap atm}. All SAPs have
non-zero positive atmosphere. Since the atmospheres now consist of pairs of
edges we have
\begin{align}
\label{eqn sap atm ratio}
\frac{\langle a_+ \rangle_{2n} }{\langle a_- \rangle_{2n+2} } &=
\frac{p_{2n+2}}{p_{2n}}
\end{align}
where $p_{2n}$ is the number of SAPs of length $2n$.

\begin{figure}[b]
  \vspace{-1ex}
\begin{center}
  \includegraphics[height=4.0cm]{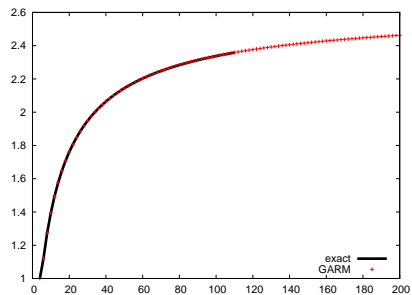}
\end{center}
  \vspace{-4ex}
\caption{A plot of $p_{n}^{1/n}$ as estimated from GARM data up to length
200 for even $n$. Exact enumeration data up to length 110 taken from
\cite{jensen2003} is shown for comparison.}
\label{fig sap2d data}
\end{figure}

In Figure~\ref{fig sap2d data} we show that data obtained by a PERM-like
implementation of GARM for SAPs agrees with exact enumeration data from
\cite{jensen2003}. We generated SAPs of a maximum of 200 edges with
approximately $4 \times 10^5$ trajectories consisting of a total of $5 \times
10^7$ samples. This took a few hours on a laptop computer.

This algorithm does not simply generalise to three dimensions with this
definition of atmospheres. If $\varphi_0$ is chosen to be a unit
square, then knotted conformations cannot be reached since inserting
atmospheric edges does not allow strand passages.

GARM can be applied to lattice bond trees. The algorithm is then closely
related to the algorithm in \cite{care2000, hsu2005}. In
this case we define the positive atmosphere by looking at all the vertices at
which an edge can be inserted to obtain a valid tree. When inserting an edge at
a given vertex, one must  be careful to consider all the possible ways of
distributing the incident branches between both ends of the new edge. One may
similarly define positive and negative atmospheres for animals. The positive
atmosphere is defined by all the ways in which an edge may be inserted at
vertices; unlike the tree case, some inserted edges will create cycles and so
are double counted as they can be inserted from either vertex. The negative
atmosphere is defined by the inverse of this process and atmospheric edges that
are not cut-edges will be double counted. This algorithm works for site-trees;
this may be easily implemented by defining positive and negative atmospheres in
terms of leaves, but more general atmospheres are possible.
  
The implementation of the GARM algorithm requires rapid calculation of the
positive and negative atmospheres. For SAWs the positive and negative
atmospheres are $O(n)$ while for SAPs they are $O(n^2)$. At present we are
able to compute the atmospheres in $O(n)$ time for SAWs and $O(n^2)$ for
SAPs. Since the atmospheres must be computed at each iteration, the time to
produce a conformation of length $n$ is $O(n^2)$ and $O(n^3)$ for SAWs and
SAPs respectively.
 
\vspace{-2ex}
\section*{Conclusions}
\vspace{-2ex}
The definitions of atmospheres above were limited to positive and negative
since they either increase or decrease the number of edges. We can generalise
this further by including the notion of neutral atmospheric moves, $a_0$, which
change the conformation without changing its size --- for example a pivot move.
At each iteration the algorithm chooses uniformly to add an edge from the
positive atmosphere or to apply a neutral atmospheric move. The probability of
obtaining a sequence $\varphi$ is
\begin{align}
  \Pr(\varphi\,|\,\varphi_0)
  &= \prod_{k=1}^{|\varphi|-1} \left( a_+(\varphi_{k-1}) + a_0(\varphi_{k-1})
\right)^{-1}.
\end{align}
and the corresponding weight is
\begin{align}
W(\varphi)
&= \prod_{k=1}^{|\varphi|-1} \frac{a_+(\varphi_{k-1})+a_0(\varphi_{k-1})}
{a_-(\varphi_k) + a_0(\varphi_k) }.
\end{align}
where $|\varphi|$ is the number of conformations in the sequence $\varphi$.
The average weight of all sequences ending in a conformation of size $n$ is
$c_n$;  the proof is as above. This addition makes it possible to sample
SAPs in three dimensions and higher since the pivot algorithm is
ergodic~\cite{madras1990}.

The algorithm can also be adapted to include Boltzmann factors (as per
\cite{seno1988, grassberger1997}) so as to compute free energies. Further
extensions such as multicanonical or flat histogram methods, such as those
developed in \cite{bachmann2003,prellberg2004} are possible. We are currently
investigating techniques to compute atmospheres more efficiently as this will
improve the convergence of the GARM algorithm.

\begin{acknowledgements}
This paper was written while visiting the Erwin Schr\"odinger Institute in
Vienna and thank them for their support. We acknowledge support from NSERC
Canada in the form of Discovery Grants. We thank Juan Alvarez, Enzo
Orlandini, Aleks Owczarek, Thomas Prellberg and Stu Whittington for
discussions and comments.
\end{acknowledgements}

\bibliography{garmbib}

\end{document}